\documentstyle[12pt]{article}
\date{}
\author{Valerii Dryuma\thanks{Work  supported in part by Grant of Programme  RFFI-Republica Moldova}\\[5mm]
{\it Institute of Mathematics and Informatics, AS RM,}\\[3mm] {\it
5 Academiei Street, 2028 Kishinev, Moldova},\\[3mm]{\it e-mail:
valery@dryuma.com; cainar@mail.md} }
\title{Around a theory  of a Walker spaces}

\begin{document}
\maketitle
\date{}
\maketitle
\begin{abstract}
\ \ \ \ The examples of a Ricci-flat  of four-dimensional spaces
with a Walker metric
\begin{equation}
ds^2=-2\Psi_i \Gamma^i_{jk}dx^idx^k+2dx^ld \Psi_l \label{dr:eq1}
\end{equation}
and their generalizations are constructed.

The properties of corresponding geodesic  equations are discussed.
\end{abstract}

\medskip
\section{A Ricci-flat of a Walker space dependent from arbitrary function}

      The properties of the of Walker metrics of the form
\begin{equation}\label{dr:eq2}
{{\it ds}}^{2}=\frac{1}{2}\left (z{\frac {\partial
}{\partial y}}\phi( x,y)+t{\frac {\partial }{\partial x}}\phi(x,y)
\right ){{\it dx}}^{2}+\left (z{\frac {
\partial }{\partial x}}\phi(x,y)+t{\frac {
\partial }{\partial y}}\phi(x,y)\right ){\it dx}\,{\it dy}+$$$$+2
\,{\it dx}\,{\it dz}+\frac{1}{2}\left (z{\frac {\partial
}{\partial y} }\phi(x,y)+t{\frac {\partial }{\partial x}}\phi(x,
y)\right ){{\it dy}}^{2}+2\,{\it dy}\,{\it dt} ,
\end{equation}
 where $\phi(x,y)$ is arbitrary function are considered.

{\bf Theorem 1.} {\it {Four dimensional space in local coordinates
$x,y,z,t$ endowed with a metric (\ref{dr:eq2}) is a Ricci-flat
\[
R_{i j}=0.
\]}
\medskip
 The proof is checked by direct calculations.

    The Riemann tensor of the metric (\ref{dr:eq2}) has the components
\[
R_{xyyt}\!=\!1/4\,{\frac {\partial ^{2}}{\partial
{x}^{2}}}\phi(x,y)-1/4 \,{\frac {\partial ^{2}}{\partial
{y}^{2}}}\phi(x,y),~
 R_{xyxz}\!=\!1/4\,{\frac {\partial
^{2}}{\partial {x}^{2}}}\phi(x,y)-1/4 \,{\frac {\partial
^{2}}{\partial {y}^{2}}}\phi(x,y),
\]
\[
R_{xyxy}\!=\!z/4\left(\frac{\partial M(x,y)}{\partial
y}\!+\!1/2\frac{\partial \phi(x,y)}{\partial x} M(x,y)
\right)\!-\!t/4\left(\frac{\partial M(x,y)}{\partial
x}\!+\!1/2\frac{\partial \phi(x,y)}{\partial y} M(x,y) \right),
\]
where
 \[
 M(x,y)=\frac{\partial^2 \phi(x,y)}{\partial
x^2}-\frac{\partial^2 \phi(x,y)}{\partial y^2},
\]
so corresponding space is a curved space.

   The geodesic of the metric are decomposed into two independent
   group of equations.

   The first one is
\[
{\frac {d^{2}x}{d{s}^{2}}}\!-\!1/4{\frac {\partial }{\partial
y}}\phi(x,y)\left ({\frac {d x}{ds}}\right )^{2}\!-\!1/2{\frac
{\partial }{\partial x}}\phi(x,y)\left ({\frac {d x}{ds} }\right
){\frac {d y}{ds}}\!-\!1/4{\frac {\partial }{
\partial y}}\phi(x,y)\left ({\frac {dy}{ds}}\right )^{2}\!=\!0,
\]
\[
{\frac {d^{2}y}{d{s}^{2}}}\!-\!1/4{\frac {\partial }{\partial
x}}\phi(x,y)\left ({\frac {dx}{ds}}\right )^{2}\!-\!1/2\,{\frac
{\partial }{\partial y}}\phi(x,y)\left ({\frac {dx}{ds} }\right
){\frac {dy}{ds}}\!-\!1/4{\frac {\partial }{
\partial x}}\phi(x,y)\left ({\frac {dy}{ds}}\right )^{2}
\!=\!0.
\]

  They do not contain the coordinates $(z,t)$ and
 are equivalent the second order ODE
\[
{\frac {d^{2}y}{d{x}^{2}}}\!+\!1/4\,\left(\frac {\partial
}{\partial y}\phi(x,y)\frac {dy}{dx}\!+\!\frac {\partial
}{\partial x}\phi(x,y)\right)\left( \left(\frac {dy}{dx}\right
)^{2}-1\right)=\!0
\]
with arbitrary function $\phi(x,y)$.

    The equation for coordinates $(z,t)$ are the linear system of
   equations with variable coefficients dependent from the coordinates
   $x(s)$ and $y(s)$ and their derivatives on the parameter  $s$.

\section{A Walker the Ricci-flat  spaces defined by Wilczynski-Tzitzeika equation}

    Wilczynski-Tzitzeika equation
\begin{equation} \label{dr:eq3}
{\frac {\partial ^{2}}{\partial x\partial
y}}\phi(x,y)=4\,{e^{2\,\phi( x,y)}}-{e^{-\phi(x,y)}}
\end{equation}
 is appeared in context of geometry of a Riemann
extensions of the spaces defined by the system of equations
\[
\ddot x-a_3(x,y)(\dot x)^2-2a_2(x,y)\dot x \dot y-a_1(x,y)(\dot
y)^2=0,
\]
\[
\ddot y+a_4(x,y)(\dot x)^2+2a_3(x,y)\dot x \dot y+a_2(x,y)(\dot
y)^2=0
\]
with arbitrary coefficients $a_i(x,y)$.

  Corresponding metric of four-dimensional Walker space is given by the expression
\begin{equation} \label{dr:eq4}
ds^2=2(za_3(x,y)-ta_4(x,y))dx^2+4(za_2(x,y)-ta_3(x,y))dxdy+$$$$+2(za_1(x,y)-ta_2(x,y))dy^2+2dxdz+2dydt
\end{equation}
 where the coefficients $a_i(x,y)$ are
\begin{equation} \label{dr:eq5}
a_1(x,y)=2\exp(\phi(x,y)),\quad a_4(x,y)=-2\exp(\phi(x,y)),$$$$
3a_2(x,y)=-\frac{\partial}{\partial y}\phi(x,y),\quad
3a_3(x,y)=\frac{\partial}{\partial x}\phi(x,y).
 \end{equation}

    A Ricci-tensor of the metric
(\ref{dr:eq4}) has the components
\[
R_{11}=2(a_{2y}-a_{1x}+2(a_1a_3-a_2^2)),\quad
R_{12}=2(a_{3y}-a_{2x}+a_1a_4-a_2a_3)
\]
\[R_{22}=2(a_{4y}-a_{3x}+2(a_2a_4-a_3^2)).
\]

{\bf Proposition1.} {\it {At the conditions (\ref{dr:eq5}) and
(\ref{dr:eq3}) a Ricci-tensor and a Riemann-tensor of the metric
(\ref{dr:eq4}) are vanish, and so the metric is a flat.}

   For  construction of a Ricci-flat but non a flat $R_{i j k l}\neq 0$
   metric defined by the solutions of the equation (\ref{dr:eq3})
    the metric conformal (\ref{dr:eq3}) with additional term
    $V(x,y)$ is introduced
\begin{equation}\label{dr:eq6}
 {{\it ds}}^{2}={\frac {\left (2\,z{\it a_3}(x,y)-2\,t{\it a_4}(x,y)
\right ){{\it dx}}^{2}+2\,\left (2\,z{\it a_2}(x,y)-2\,t{\it
a_3}(x,y)+2 \,V(x,y)\right ){\it dx}\,{\it dy}}{\left
(U(x,y)\right )^{2}}}+$$$$+{ \frac {2\,{\it dx}\,{\it dz}+\left
(2\,z{\it a_1}(x,y)-2\,t{\it a_2}(x,y )\right ){{\it
dy}}^{2}+2\,{\it dy}\,{\it dt}}{\left (U(x,y)\right )^{ 2}}}.
\end{equation}

{\bf Theorem 2.} {\it {At the conditions (\ref{dr:eq5}) and
(\ref{dr:eq3}) the metric (\ref{dr:eq6}) is a Ricci-flat
\[
R_{i j}=0
\]}
(but non a flat $R_{i j k l}\neq 0$), if the function $U(x,y)$ is
solution of compatible system of equations
\[{\frac {\partial ^{2}}{\partial {y}^{2}}}U(x,y)=\]\[=\left (1/3\,{\frac {
\partial ^{2}}{\partial {y}^{2}}}\phi(x,y)+2/9\,\left ({\frac {
\partial }{\partial y}}\phi(x,y)\right )^{2}+2/3\,\left ({\frac {
\partial }{\partial x}}\phi(x,y)\right ){e^{\phi(x,y)}}\right
)U(x,y)-\]\[- 1/3\,\left ({\frac {\partial }{\partial
y}}\phi(x,y)\right ){\frac {
\partial }{\partial y}}U(x,y)-2\,{e^{\phi(x,y)}}{\frac {\partial }{
\partial x}}U(x,y),
\]
\[
{\frac {\partial ^{2}}{\partial x\partial y}}U(x,y)=\]\[=\left
(2/3\,{e^{- \phi(x,y)}}-1/9\,\left ({\frac {\partial }{\partial
x}}\phi(x,y) \right ){\frac {\partial }{\partial
y}}\phi(x,y)+4/3\,{e^{2\,\phi(x,y) }}\right
)U(x,y)+\]\[+1/3\,\left ({\frac {\partial }{\partial y}}\phi(x,y)
\right ){\frac {\partial }{\partial x}}U(x,y)+1/3\,\left ({\frac {
\partial }{\partial x}}\phi(x,y)\right ){\frac {\partial }{\partial y}
}U(x,y),
\]
\[
{\frac {\partial ^{2}}{\partial {x}^{2}}}U(x,y)=\]\[=\left
(1/3\,{\frac {
\partial ^{2}}{\partial {x}^{2}}}\phi(x,y)+2/9\,\left ({\frac {
\partial }{\partial x}}\phi(x,y)\right )^{2}+2/3\,\left ({\frac {
\partial }{\partial y}}\phi(x,y)\right ){e^{\phi(x,y)}}\right
)U(x,y)-\]\[- 1/3\,\left ({\frac {\partial }{\partial
x}}\phi(x,y)\right ){\frac {
\partial }{\partial x}}U(x,y)-2\,{e^{\phi(x,y)}}{\frac {\partial }{
\partial y}}U(x,y).
\]}
The function $V(x,y)$ in so doing is arbitrary.
\medskip

The proof of theorem is checked by direct calculations.

 {\small
\centerline{\bf References:}

\smallskip
\noindent 1. A.G. Walker, {\it Canonical form of a Riemannian
spaces with a parallel field of null planes}. Q. J. Math.
Oxford(2) {\bf 1} (1950), p. 69--79.

\smallskip
\noindent 2. V. Dryuma, {\it The Riemann and Einsten-Weyl
geometries in theory of differential equations, their applications
and all that}. A.B. Shabat et all.(eds.), New Trends in
Integrability and Partial Solvability, Kluwer Academic Publishers,
Printed in the Netherlands , 2004,  p.115--156.}

\smallskip
\noindent 3.
 V. Dryuma, {\it Toward a theory of spaces of constant curvature. Theoretical and Mathematical Physics,
 v. 146(1): 2006, p.35-45, ArXiv: math. DG/0505376, 18 May 2005, p.1-12.}

\smallskip
\noindent 4.
 V. Dryuma, {\it 10-Dim Einstein spaces made up on
basis 6-dim Ricci-flat spaces and 4-dim Einstein spaces, ArXiv:
gr-qc/06060403 v 1, 9 Jun, 2006, p.1--8.}

\smallskip
\noindent 5.
 V. Dryuma, {\it Four-dim Einstein spaces on a six-dim Ricci-flat base space,
ArXiv: gr-qc/0601051 v 1, 12 Jan, 2006, p.1--9.}
\end{document}